\documentclass{article}

\newif\ifarxiv
\arxivtrue

\usepackage{graphicx}
\usepackage{amsmath,amssymb,amsthm,bm}
\usepackage[a4paper]{geometry}
\usepackage{algorithm,algorithmicx,algpseudocode}
\newcommand*{\SComment}[1]{\Comment{{\scriptsize #1}}} 
\usepackage[colorlinks=false,pdfborder={0 0 0}]{hyperref}
\usepackage{authblk}
\usepackage{color}
\usepackage{comment}
\usepackage{xspace}
\usepackage{subfigure}

\graphicspath{{fig/}}
\DeclareGraphicsExtensions{.mps,.pdf,.eps,.png}

\newenvironment{keywords}{\medskip\textbf{Keywords:}}{}
\newenvironment{AMS}{\medskip\textbf{AMS subject classifications (2020).}}{}

\theoremstyle{plain}

\newcommand*{\trans}{^{\top}}

\newcommand*{\itrans}{^{-\top}}

\newcommand{\bmat}[1]{\begin{bmatrix}#1\end{bmatrix}}

\newcommand*{\norm}[1]{\Vert#1\rVert}

\def\adots{\mathinner{\mkern2mu\raise1pt\hbox{.}\mkern2mu
    \raise4pt\hbox{.}\mkern2mu\raise7pt\hbox{.}\mkern1mu}}

\newcommand{\sigmin}{\sigma_{\min}}

\newcommand{\vQ}{\bm{Q}}

\newcommand{\vU}{\bm{U}}

\newcommand{\vX}{\bm{X}}

\newcommand{\vZ}{\bm{Z}}

\newcommand{\RR}{\mathcal{R}}

\renewcommand{\SS}{\mathcal{S}}

\newcommand{\YY}{\mathcal{Y}}

\newcommand{\bQQ}{\bm{\mathcal{Q}}}

\newcommand{\bXX}{\bm{\mathcal{X}}}

\newcommand{\BCGSIRO}{\texttt{BCGSI+}\xspace}	
\newcommand{\BCGSIROAPIPIROone}{\texttt{BCGSI+P-1S-2S}\xspace}	
\newcommand{\BCGSIROAone}{\texttt{BCGSI+1s}\xspace}	
\newcommand{\BCGSPIPIRO}{\hyperref[alg:BCGSPIPIRO]{\texttt{BCGS-PIPI+}}\xspace}	
\newcommand{\BCGSPIPIROone}{\hyperref[alg:BCGSPIPIRO1S]{\texttt{BCGSI+P-1S}}\xspace}
\newcommand{\BCGSPIPIROtwo}{\hyperref[alg:BCGSPIPIRO2S]{\texttt{BCGSI+P-2S}}\xspace}

\newcommand{\TSQR}{\texttt{TSQR}\xspace}	
\newcommand{\CholQR}{\texttt{CholQR}\xspace}	

\newcommand*{\chol}{\texttt{chol}\xspace}

\newcommand*{\macheps}{\bm u}

\newcommand*{\bigO}{O}

\definecolor{plotred}{RGB}{216.7500, 82.8750, 24.9900}
\newcommand{\diff}[1]{\textcolor{black}{#1}}

\title{The Performance of Low-Synchronization Variants of Reorthogonalized Block Classical Gram--Schmidt}
\author[1]{Erin Carson}
\author[1]{Yuxin Ma}
\affil[1]{Department of Numerical Mathematics, Faculty of Mathematics and Physics, Charles University, Sokolovsk\'{a} 49/83, 186 75 Praha 8, Czechia

\texttt{Email: carson@karlin.mff.cuni.cz, yuxin.ma@matfyz.cuni.cz}}

\begin{document}
\maketitle

\begin{abstract}
Numerous applications, such as Krylov subspace solvers, make extensive use of the block classical Gram-Schmidt (BCGS) algorithm and its reorthogonalized variants for orthogonalizing a set of vectors.
For large-scale problems in distributed memory settings, the communication cost, particularly the global synchronization cost, is a major performance bottleneck.
In recent years, many low-synchronization BCGS variants have been proposed in an effort to reduce the number of synchronization points.
The work [E. Carson, Y. Ma, arXiv preprint 2411.07077] recently proposed stable one-synchronization and two-synchronization variants of BCGS, i.e., \BCGSPIPIROone and \BCGSPIPIROtwo.
In this work, we evaluate the performance of \BCGSPIPIROone and \BCGSPIPIROtwo on a distributed memory system compared to other well-known low-synchronization BCGS variants.
In comparison to the classical reorthogonalized BCGS algorithm (\BCGSIRO), numerical experiments demonstrate that \BCGSPIPIROone and \BCGSPIPIROtwo can achieve up to \(4\times\) and \(2\times\) speedups, respectively, and perform similarly to other (less stable) one-synchronization and two-synchronization variants.
\BCGSPIPIROone and \BCGSPIPIROtwo are therefore recommended as the best choice in practice for computing an economic QR factorization on distributed memory systems due to their superior stability when compared to other variants with the same synchronization cost.


\end{abstract}




\section{Introduction}
\label{sec:introduction}
In this work, we consider employing the classical Gram--Schmidt (CGS) algorithm to compute the economic QR factorization of an \(n\times m\) matrix \(\bXX\), that is,
$\bXX = \bQQ \RR$,
where \(\bQQ \in \mathbb{R}^{m\times n}\) has orthogonal columns and \(\RR\in \mathbb{R}^{n \times n}\) is upper triangular.

On distributed systems, reducing the synchronization points, which are defined as global communication involving all processors such as \texttt{MPI\_Allreduce}, has become increasingly crucial for performance.
Blocking, which reduces synchronization points, is therefore a compelling strategy; see the survey~\cite{CLRT2022}.
Consequently, much attention has been paid to the block classical Gram-Schmidt (BCGS) algorithm and its low-synchronization (low-sync) variants.
The traditional BCGS algorithm needs only two synchronization points per iteration (i.e., per block column) but has an instability issue, as its loss of orthogonality \(\norm{\bQQ\trans \bQQ - I}\) is not \(\bigO(\macheps)\) as demonstrated by~\cite{CLMO2024-ls}, where $\macheps$ denotes the unit roundoff.
A reorthogonalized variant, \BCGSIRO, was proposed, which achieves \(\big(\macheps)\)  at the cost of adding two additional synchronization points per iteration.

The low-sync versions are essentially derived from \BCGSIRO in order to decrease the added synchronization points.
Numerous studies~\cite{CLMO2024-ls,CLMO2024-P,CM2024-onesync} have proposed various two-sync BCGS variants, like \BCGSPIPIRO and \BCGSPIPIROtwo, and one-sync BCGS variants, like \BCGSPIPIROone and \BCGSIROAone, and have examined the theoretical and experimental characteristics of loss of orthogonality in recent years.
These works primarily focus on the loss of orthogonality property rather than performance.

In this work, we evaluate how well these various low-sync BCGS variants perform on distributed memory systems.
We summarize the most common low-sync BCGS variants in Section~\ref{sec:algo}.
The MPI parallelization strategy for these algorithms is then provided in Section~\ref{sec:mpiparal}.
Section~\ref{sec:experiments} compares \BCGSIRO on a distributed memory system with all the low-sync BCGS variants described in Section~\ref{sec:algo}.

For simplicity, we take \(m = qs\) and let \(\bXX \in \mathbb{R}^{n\times qs}\) be composed of \(q\) block vectors (or block columns) so that \(\bXX = \bmat{\vX_1& \dotsi & \vX_q}\) where \(\vX_k \in \mathbb{R}^{n\times s}\) for any \(k = 1, 2, \dotsc, q\).
We then introduce some notation used throughout the paper before proceeding.
Uppercase Roman scripts, for example, \(\bXX\) and \(\bQQ\), denote the matrices containing \(q\) block vectors.
Each block vector is denoted by uppercase Roman letters \(\vX_k\) and \(\vQ_k\), i.e., \(\bXX_k = \bmat{\vX_1& \vX_2& \dotsi & \vX_k}\) and \(\bQQ_k = \bmat{\vQ_1& \vQ_2& \dotsi & \vQ_k}\).
For square matrices, we use uppercase Roman scripts, for example, \(\RR\), \(\SS\), and \(\YY\), to represent \(qs \times qs\) matrices.
MATLAB indexing is used to denote submatrices.
For example, $\RR_{1:k-1,k} = \bmat{R_{1,k}\trans & R_{2,k}\trans & \cdots & R_{k-1,k}\trans}\trans$. 
In addition, \(\RR_{kk} = \RR_{1:k, 1:k}\), \(\SS_{kk} = \SS_{1:k, 1:k}\), and \(\YY_{kk} = \YY_{1:k, 1:k}\).
We also use \(\kappa(\bXX) = \norm{\bXX}/\sigmin(\bXX)\) to represent the 2-norm condition number, where \(\norm{\cdot}\) is the \(2\)-norm and \(\sigmin(\bXX)\) is the smallest singular value of \(\bXX\).
The term intraorthogonalization means the orthogonalization or the economic QR factorization of a block vector $\vX$.
Various algorithms such as TSQR (\TSQR), CGS, or Cholesky QR (\CholQR), described in~\cite{CLRT2022}, can be applied as intraorthogonalization routines.

\section{Low-synchronization BCGS algorithms}
\label{sec:algo}
In this section, we provide an overview of two stable low-synchronization variants of BCGS, i.e., \BCGSPIPIROone and \BCGSPIPIROtwo, and other low-synchronization variants including \BCGSPIPIRO, \BCGSIROAone. We provide a summary of all these BCGS variants in Table~\ref{table:methods} so that the properties of these approaches can be readily compared.


\subsection{Two-sync BCGS: \BCGSPIPIRO}
In \cite{CLMO2024-P}, the authors study a reorthogonalized variant of BCGS using the Pythagorean inner product, i.e., \BCGSPIPIRO, summarized in Algorithm~\ref{alg:BCGSPIPIRO}.
To reduce the number of synchronization points needed for \TSQR, \BCGSPIPIRO uses the Pythagorean-based Cholesky QR algorithm instead of the unconditionally stable QR algorithm, i.e., \TSQR.
This means that only two synchronization points are needed for each iteration.
This variant has been demonstrated to have \(\bigO(\macheps)\) loss of orthogonality; however, because \chol is used in Line~\ref{line:bcgspipiro:Sdiag}, it necessitates the assumption \(\bigO(\macheps)\kappa^2(\bXX)\leq 1\).

\begin{algorithm}[htbp!]
\footnotesize
    \caption{$[\bQQ, \RR] = \BCGSPIPIRO(\bXX)$ \label{alg:BCGSPIPIRO}}
    \begin{algorithmic}[1]
        \State{$[\vQ_1, R_{11}] = \TSQR(\vX_1)$}
        \For{$k = 1, \ldots, q-1$}
            \State{$\bmat{\SS_{1:k,k+1} \\ T_{k+1}} = \bmat{\bQQ_{k} & \vX_{k+1}}\trans \vX_{k+1}$} \SComment{First synchronization} \label{line:bcggspipiro:SS}
            \State{$S_{k+1,k+1} = \chol\bigl( T_{k+1} - \SS_{1:k,k+1}\trans \SS_{1:k,k+1} \bigr)$} \label{line:bcgspipiro:Sdiag}
            \State{$\vU_{k+1} = (\vX_{k+1} - \bQQ_{k} \SS_{1:k,k+1}) S_{k+1,k+1}^{-1}$} \label{line:bcggspipiro:Uk}
            \State{$\bmat{\YY_{1:k,k+1} \\ \Omega_{k+1}} = \bmat{\bQQ_{k} & \vU_{k+1}}\trans \vU_{k+1}$} \SComment{Second synchronization} \label{line:bcggspipiro:YY-Omega}
            \State $Y_{k+1,k+1} = \chol\bigl( \Omega_{k+1} - \YY_{1:k,k+1}\trans \YY_{1:k,k+1} \bigr), \quad
            \vQ_{k+1} = (\vU_{k+1} - \bQQ_{k} \YY_{1:k,k+1}) Y_{k+1,k+1}^{-1}$ 
            \State $\RR_{1:k,k+1} = \SS_{1:k,k+1} + \YY_{1:k,k+1} S_{k+1,k+1}, \quad R_{k+1,k+1} = Y_{k+1,k+1} S_{k+1,k+1}$ \label{line:bcgspipiro:Rdiag}
        \EndFor
        \State \Return{$\bQQ = [\vQ_1, \ldots, \vQ_q]$, $\RR = (R_{ij})$}
    \end{algorithmic}
\end{algorithm}

\subsection{One-sync BCGS: \BCGSPIPIROone and \BCGSIROAone}
Algorithm~\ref{alg:BCGSPIPIRO} can have fewer synchronization points by using the delayed normalization technique~\cite{YTHBSE2020}.
Specifically, by splitting the operation \(\bQQ_{k}\trans \vX_{k+1}\) in Line~\ref{line:bcggspipiro:SS} of Algorithm~\ref{alg:BCGSPIPIRO} into two parts based on \(\bQQ_{k}\trans \vX_{k+1} = \bmat{\bQQ_{k-1}\trans \vX_{k+1}& \vQ_k\trans \vX_{k+1}}\),
and then using 
\begin{equation} \label{eq:QkXk+1}
    \begin{split}
        \vQ_k\trans \vX_{k+1} &= Y_{kk}\itrans (\underbrace{\vU_k\trans \vX_{k+1}}_{=: P_k} - \YY_{1:k-1,k}\trans \underbrace{\bQQ_{k-1}\trans \vX_{k+1}}_{=: \vZ_{k-1}})
        = Y_{kk}\itrans (P_k - \YY_{1:k-1,k}\trans \vZ_{k-1})
    \end{split}
\end{equation}
to compute the second part \(\vQ_k\trans \vX_{k+1}\), \BCGSPIPIRO, i.e., Algorithm~\ref{alg:BCGSPIPIRO}, can be reorganized as Algorithm~\ref{alg:BCGSPIPIRO1S}.
This one-sync variant, called \BCGSPIPIROone, is proposed by~\cite{CM2024-onesync}. It only requires one sync point per iteration, with the same loss of orthogonality property as \BCGSPIPIRO.

A different one-sync variant, \BCGSIROAone, also known as \texttt{BCGSI+LS}, is introduced in~\cite{CLMO2024-ls, YTHBSE2020}.
It is based on the delayed normalization technique, but it does not use the Pythagorean-based Cholesky QR algorithm for the first intraorthogonalization.
Compared to \BCGSPIPIROone, there is one less step per iteration, i.e., Line~\ref{line:bcgspipiro1s:chol-1} of Algorithm~\ref{alg:BCGSPIPIRO1S}, at the cost of instability.
This variant only has \(\bigO(\macheps)\kappa^2(\bXX)\) loss of orthogonality with the assumption \(\bigO(\macheps)\kappa^3(\bXX)\leq 1\) as shown in~\cite{CLMO2024-ls}.

\begin{algorithm}[!tb]
\footnotesize
    \caption{$[\bQQ, \RR] = \BCGSPIPIROone(\bXX)$ \label{alg:BCGSPIPIRO1S}}
    \begin{algorithmic}[1]
        \State $[\vQ_1, R_{11}] = \TSQR(\vX_1),
        \quad \bmat{S_{12} \\ T_2} = \bmat{\vQ_1& \vX_2}\trans \vX_2, \quad S_{22} = \chol (T_2 - S_{12}\trans S_{12}),
        \label{line:bcgspipiro1s:S22}
        \quad \vU_2 = (\vX_2 - \vQ_1 S_{12}) S_{22}^{-1}$
        \State {$\begin{bmatrix} \YY_{12} & \vZ_{1} \\ \Omega_2 & P_2 \end{bmatrix} = \begin{bmatrix}\vQ_{1} & \vU_2 \end{bmatrix}\trans \begin{bmatrix}\vU_2 & \vX_{3} \end{bmatrix}$} \SComment{Synchronization}
        \State $T_{3} = \vX_{3}\trans \vX_{3}, \quad 
        Y_{22} = \chol(\Omega_2 - \YY_{12}\trans \YY_{12}), \quad
        \vQ_2 = (\vU_2 - \vQ_{1} \YY_{12}) Y_{22}^{-1}, \quad 
        \RR_{12} = \SS_{12} + \YY_{12} \diff{S_{22}}, \quad
        R_{22} = Y_{22} \diff{S_{22}}$
        \For{$k = 2, \ldots, q-1$}
            \State $\SS_{1:k,k+1} =
                \begin{bmatrix} \vZ_{k-1} \\
                Y_{kk}\itrans \left( P_k - \YY_{1:k-1,k}\trans \vZ_{k-1} \right) \end{bmatrix}, \quad 
            S_{k+1, k+1} = \chol (T_{k+1} - \SS_{1:k,k+1}\trans \SS_{1:k,k+1})$ \label{line:bcgspipiro1s:chol-1}
            \State \diff{$\vU_{k+1} = (\vX_{k+1} - \bQQ_k \SS_{1:k,k+1}) S_{k+1, k+1}^{-1}$} \label{line:1s:Uk}
            \If{$k < q-1$}
                \State $\begin{bmatrix} \YY_{1:k,k+1} & \vZ_{k} \\ \Omega_{k+1} & P_{k+1} \end{bmatrix} =
                \begin{bmatrix}\bQQ_{k} & \vU_{k+1} \end{bmatrix}\trans \begin{bmatrix}\vU_{k+1} & \vX_{k+2} \end{bmatrix}$ \SComment{Synchronization} \label{line:bcgspipiro1s:sync}
                \Statex $\qquad\quad T_{k+2} = \vX_{k+2}\trans \vX_{k+2}$
            \Else
                \State $\begin{bmatrix} \YY_{1:k,k+1} \\ \Omega_{k+1}\end{bmatrix} =
                \begin{bmatrix}\bQQ_{k} & \vU_{k+1} \end{bmatrix}\trans \vU_{k+1}$ 
            \EndIf 
            \State $Y_{k+1,k+1} = \chol(\Omega_{k+1} - \YY_{1:k,k+1}\trans \YY_{1:k,k+1}), \quad \vQ_{k+1} = (\vU_{k+1} - \bQQ_{k} \YY_{1:k,k+1}) Y_{k+1,k+1}^{-1}$ \label{line:Qk}
            \State $\RR_{1:k,k+1} = \SS_{1:k,k+1} + \YY_{1:k,k+1} \diff{S_{k+1,k+1}}, \quad R_{k+1,k+1} = Y_{k+1,k+1} \diff{S_{k+1,k+1}}$
        \EndFor
        \State \Return{$\bQQ = [\vQ_1, \ldots, \vQ_q]$, $\RR = (R_{ij})$}
    \end{algorithmic}
\end{algorithm}

\subsection{Two-sync BCGS: \BCGSPIPIROtwo}
The requirement \(\bigO(\macheps)\kappa^2(\bXX)\leq 1\) for loss of orthogonality is too stringent for some applications, such as Krylov subspace methods~\cite{CM2024}.
Thus, the authors of~\cite{CM2024-onesync} propose \BCGSPIPIROtwo, which uses an unconditional stable algorithm \TSQR for the first intraorthogonalization instead of \chol in both Lines~\ref{line:bcgspipiro1s:S22} and~\ref{line:bcgspipiro1s:chol-1} of Algorithm~\ref{alg:BCGSPIPIRO1S}, to relax the condition to \(\bigO(\macheps)\kappa(\bXX)\leq 1\) while maintaining \(\bigO(\macheps)\) loss of orthogonality at the cost of adding one more synchronization point per iteration.

\begin{table}[!tb]
\footnotesize
\centering
\caption{The properties of \BCGSIRO and various low-sync BCGS variants. In this table, we use LOO to denote loss of orthogonality \(\norm{\bQQ\trans\bQQ-I}\).}
\label{table:methods}
\begin{tabular}{l|lll}
\hline
Methods             & Sync points & LOO & Assumption for LOO \\ \hline
\texttt{BCGSI+P-1s} & \(q\)    & \(\bigO(\macheps)\)     & \(\bigO(\macheps)\kappa^2(\bXX)\leq 1\)     \\
\texttt{BCGSI+1s}   & \(q\)     & \(\bigO(\macheps)\kappa^2(\bXX)\)  & \(\bigO(\macheps)\kappa^3(\bXX)\leq 1\)   \\
\texttt{BCGSI+P-2s} & \(2q\)     & \(\bigO(\macheps)\)     & \(\bigO(\macheps)\kappa(\bXX)\leq 1\)\\
\texttt{BCGSPIPI+}  & \(2q\)     & \(\bigO(\macheps)\)    & \(\bigO(\macheps)\kappa^2(\bXX)\leq 1\)   \\
\texttt{BCGSI+}     & \(4q\)     & \(\bigO(\macheps)\)    & \(\bigO(\macheps)\kappa(\bXX)\leq 1\)     \\ \hline
\end{tabular}
\end{table}

\section{MPI parallelization}
\label{sec:mpiparal}
This section presents our MPI parallelization approach for these BCGS low-synchronization variations.
For simplicity, we demonstrate our MPI parallelization approach using \BCGSPIPIROone, i.e., Algorithm~\ref{alg:BCGSPIPIRO1S}, as an example.
The remaining variations are comparable.

The given matrix \(\bXX\) and the result \(\bQQ\) are distributed using a 1D rowwise distribution, assuming that there are \(p\) MPI processes and that \(m\)-by-\(m\) and \(\lceil n/p\rceil\)-by-\(m\) matrices can be stored in the memory of each process.
For example, the given matrix \(\bXX\) is divided into \(p\) non-overlapping submatrices \(\bXX_{1,1:q}\), \(\dotsc\), \(\bXX_{p,1:q}\), i.e.,
$\bXX = \bmat{\bXX_{1,1:q}\trans &\bXX_{2,1:q}\trans & \cdots & \bXX_{p,1:q}\trans}\trans$,
and each MPI process owns one submatrix.
All of the processes own the small output matrix \(\RR\).
The 1D rowwise distribution is also used to store the intermediate variable \(\vU\) in Algorithm~\ref{alg:BCGSPIPIRO1S}, while each process owns the other small intermediate variables.
The multiplication of two tall-thin matrices, such as \(\begin{bmatrix}\bQQ_{k} & \vU_{k+1} \end{bmatrix}\trans \begin{bmatrix}\vU_{k+1} & \vX_{k+2} \end{bmatrix}\), is the only fundamental operation involving global communication that is included in \BCGSPIPIROone.
Each process determines the multiplication of its own components before obtaining the final result through a single global synchronization.
\section{Numerical Experiments}
\label{sec:experiments}
This section presents parallel numerical experiments on a distributed memory system for \(\BCGSIRO\), \(\BCGSPIPIROone\), \(\BCGSPIPIROtwo\), \(\BCGSIROAPIPIROone\), \(\BCGSPIPIRO\), and \(\BCGSIROAone\), discussed in Section~\ref{sec:algo}.
Our implementations use the \TSQR routine from the Trilinos C++ library~\cite{trilinos2005}.
The experiments were performed on the cluster named Karolina%
\footnote{https://docs.it4i.cz}
with 720 nodes, each equipped with one 64-core CPU.
We utilized up to 128 nodes, assigning one process
per node as the mode of parallelization.
We only concentrate on the distributed memory performance and do not test the accuracy of these BCGS algorithms, as this has already been examined in~\cite{CM2024-onesync}.
Therefore, we use random matrices as test matrices for simplicity.

\subsection{Tests for different block size \(s\)}
Table~\ref{table:s} displays the speedup of different BCGS variants with different \(s\) compared to \texttt{CGSI+}, i.e., \BCGSIRO with \(s=1\), for the test matrix with the fixed size \(10^6\times 400\) employing 64 nodes.
As the performance is nearly entirely dependent on synchronization, a larger block size \(s\) can result in a better speedup because it also means fewer synchronization points.
One-sync and two-sync BCGS variants achieve, respectively, up to approximately \(4\times\) and \(2\times\) for each \(s\) in comparison to \BCGSIRO.

\begin{table}[!tb]
\footnotesize
\centering
\caption{Speedup of different BCGS variants with different \(s\) compared to \texttt{CGSI+} with \(s=1\) using \(64\) nodes: \(n=10^6\) rows, \(m=400\) columns, varying block size \(s\).}
\label{table:s}
\begin{tabular}{l|lllll}
\hline
Methods                              & $s = 1$ & $s = 2$ & $s = 4$ & $s = 8$ & $s = 16$ \\ \hline
\texttt{BCGSI+P-2s} & 1.3$\times$    & 2.6$\times$     & \hphantom{0}5.2$\times$     & 10.2$\times$    & 15.9$\times$     \\
\texttt{BCGSI+P-1s} & 2.8$\times$     & 5.4$\times$     & 10.8$\times$    & 20.0$\times$    & 37.5$\times$     \\
\texttt{BCGSI+1s}   & 2.9$\times$     & 5.8$\times$     & 10.6$\times$    & 21.2$\times$    & 37.8$\times$    \\
\texttt{BCGSPIPI+}  & 1.6$\times$    & 3.0$\times$    & \hphantom{0}5.9$\times$    & 11.7$\times$   & 22.7$\times$    \\
\texttt{BCGSI+}     & 1.0$\times$    & 1.4$\times$    & \hphantom{0}2.7$\times$    & \hphantom{0}5.4$\times$    & \hphantom{0}8.8$\times$     \\ \hline
\end{tabular}
\end{table}

\subsection{Tests for different number of rows and columns}
\label{subsec:experiments:row-col}
\begin{figure}[!tb]
\centering
\includegraphics[width=.8\textwidth]{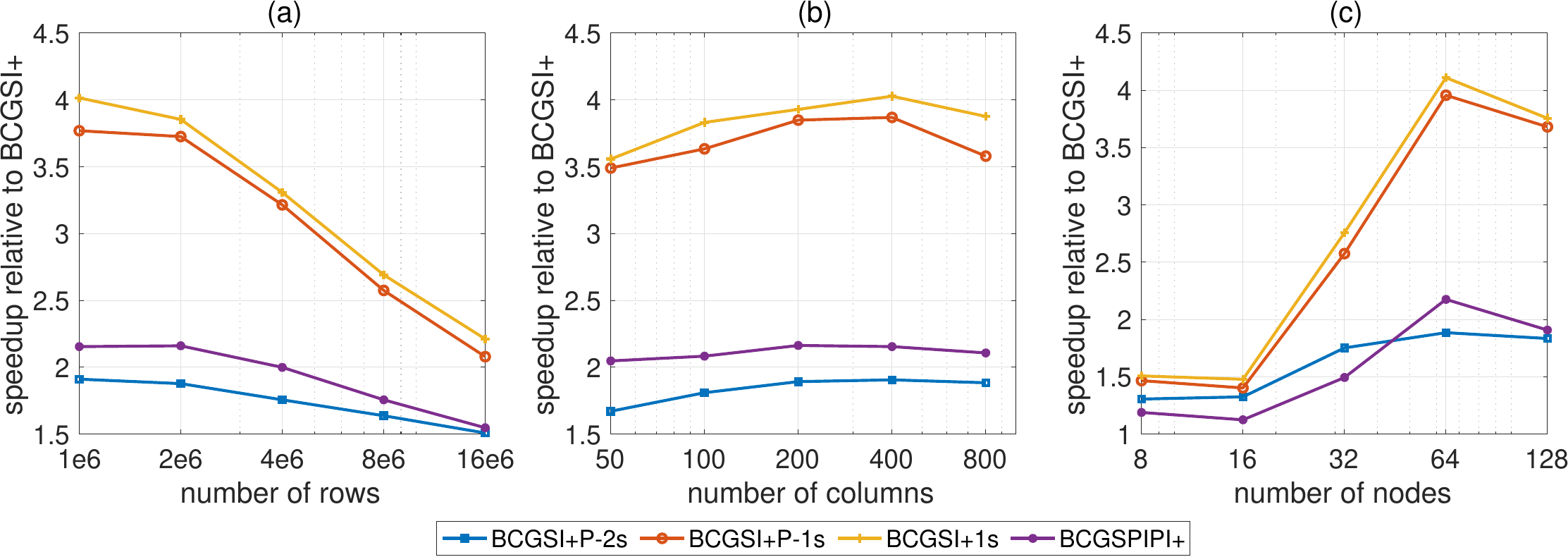}
\caption{Speedup of different BCGS variants compared to \BCGSIRO: In Figure (a), the setup consists of \(64\) nodes with a block size of \(s=4\), utilizing \(m=400\) columns and altering the row count \(n\).
Meanwhile, Figure (b) also uses \(64\) nodes and a block size of \(s=4\), but instead, it operates with \(n=10^6\) rows and adjusts the column count \(m\).
Figure (c) utilizes \(n=10^6\) rows and \(m=400\) columns, with a block size of \(s=4\), and varies the number of nodes.}
\label{fig:row}
\end{figure}

Figures~\ref{fig:row} (a) and~(b) present the speedup of different BCGS variants compared to \BCGSIRO also using \(64\) nodes with fixed block size \(s=4\).
In Figure~\ref{fig:row} (a), we fix the number of columns \(m=400\) and vary the number of rows \(n\).
As the number of rows increases, each process needs to process more calculations; thus, the proportion of run time spent on synchronization is reduced.
Therefore, it can be seen that all speedups are reduced by increasing the number of rows.

In Figure~\ref{fig:row} (b), we vary \(m\) with fixed \(n=10^6\).
Note that varying the number of columns does not change the amount of data each process handles and only increases the number of iterations that the BCGS algorithms perform.
Therefore, it is reasonable that varying the number of columns only has a slight effect on the performance comparison between these BCGS variants.

\subsection{Strong-scaling tests}
\label{subsec:experiments:strong-scal}
As the cost of synchronization increases with the number of nodes, Figure~\ref{fig:row} (c) illustrates that for the strong-scaling case, the speedups increase in magnitude.
In situations involving more than 64 nodes, the cost of synchronization nearly always dominates the cost of orthogonalizing a \(10^6\times 400\) matrix. 
Low-synchronization BCGS variants thus attain the highest speedups.


\subsection{Weak-scaling tests}
\begin{figure}[!tb]
\centering
\includegraphics[width=.7\textwidth]{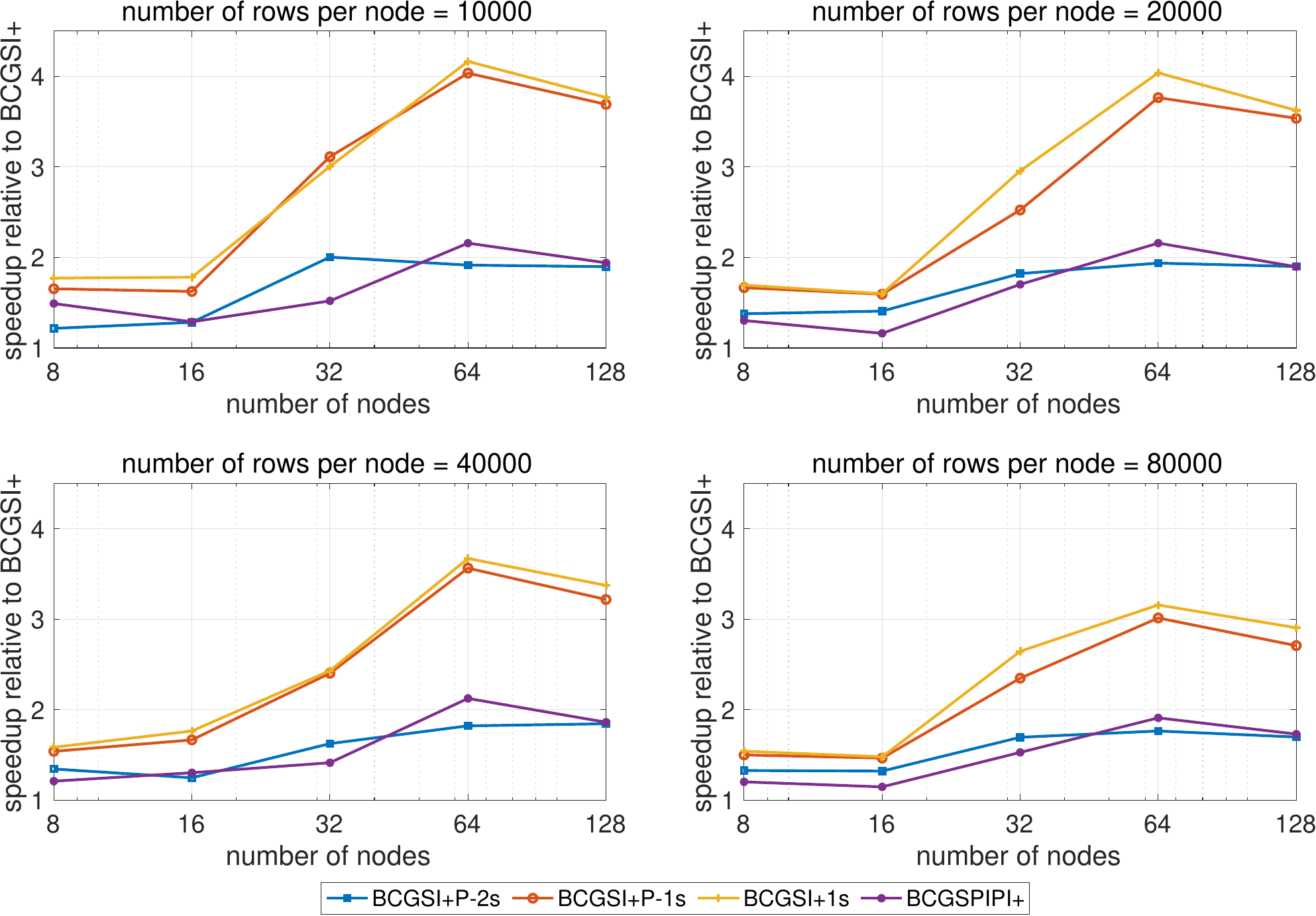}
\caption{Speedup of different BCGS variants compared to \texttt{BCGSI+}: \(m=400\) columns, block size \(s=4\), varying number of nodes with fixed number of rows per node.}
\label{fig:weakscal}
\end{figure}

To demonstrate the performance of the weak-scaling case, this subsection varies the number of nodes while maintaining the fixed columns \(m=400\) and the fixed number of rows per node.
We do not need to illustrate the results of the weak-scaling case shown in Figure~\ref{fig:weakscal} since they are similar to the results of Sections~\ref{subsec:experiments:row-col} and~\ref{subsec:experiments:strong-scal}.
We notice that there are instances where the speedups marginally decline as the number of nodes increases from 8 to 16.
This occurs because the run times for scenarios using either 8 or 16 nodes are quite short, leading to unstable speedups, despite us averaging over 10 run per case.

\medskip
In conclusion, the two one-sync variants have comparable performance and can both produce up to \(4\times\) speedups when compared to \BCGSIRO.
Furthermore, the two two-sync variants have comparable performance and can both produce up to \(2\times\) speedups when compared to \BCGSIRO.
Both theoretically and experimentally, however, \BCGSPIPIROone and \BCGSPIPIROtwo have a better loss of orthogonality property than \BCGSIROAone and \BCGSPIPIRO, respectively, as demonstrated in~\cite{CM2024-onesync}.
Thus \BCGSPIPIROone or \BCGSPIPIROtwo are preferred in practice, depending on the required constraint on the condition number of the input. 
\section{Conclusions}
\label{sec:conclusions}
In this work, we summarize all of the well-known low-synchronization BCGS variants, present the MPI parallelization strategy, and contrast their performance with that of \BCGSIRO on a distributed memory system.
Our numerical results first show that performance can be improved by reducing the synchronization points with a larger block size \(s\).
The numerical results also demonstrate that BCGS variants with the same number of synchronization points perform similarly, with one-sync and two-sync variants achieving up to \(4\times\) and \(2\times\) speedups, respectively, in comparison to \BCGSIRO.
We recommend \BCGSPIPIROone and \BCGSPIPIROtwo in practice since, as established in~\cite{CM2024-onesync}, they have a better loss of orthogonality property than \BCGSIROAone and \BCGSPIPIRO, respectively.
\section*{Acknowledgments}
Both authors are supported by the European Union (ERC, inEXASCALE, 101075632). Views and opinions expressed are those of the authors only and do not necessarily reflect those of the European Union or the European Research Council. Neither the European Union nor the granting authority can be held responsible for them. Both authors additionally acknowledge support from the Charles University Research Centre program No. UNCE/24/SCI/005.

\bibliographystyle{abbrvurl}
\bibliography{mybib}

\begin{thebibliography}{1}

\bibitem{CLMO2024-ls}
E.~Carson, K.~Lund, Y.~Ma, and E.~Oktay.
\newblock On the loss of orthogonality in low-synchronization variants of reorthogonalized block classical {{Gram--Schmidt}}, 2024.
\newblock \href {https://arxiv.org/abs/2408.10109} {\path{arXiv:2408.10109}}.

\bibitem{CLMO2024-P}
E.~Carson, K.~Lund, Y.~Ma, and E.~Oktay.
\newblock Reorthogonalized {P}ythagorean variants of block classical {Gram--Schmidt}.
\newblock {\em SIAM J. Matrix Anal.\ Appl.}, 46(1):310--340, 2025.
\newblock \href {https://doi.org/10.1137/24M1658723} {\path{doi:10.1137/24M1658723}}.

\bibitem{CLRT2022}
E.~Carson, K.~Lund, M.~Rozlo{\v z}n{\'i}k, and S.~Thomas.
\newblock Block {{Gram--Schmidt}} algorithms and their stability properties.
\newblock {\em Linear Algebra Appl.}, 638(20):150--195, 2022.
\newblock \href {https://doi.org/10.1016/j.laa.2021.12.017} {\path{doi:10.1016/j.laa.2021.12.017}}.

\bibitem{CM2024}
E.~Carson and Y.~Ma.
\newblock On the backward stability of s-step {GMRES}, 2024.
\newblock \href {https://arxiv.org/abs/2409.03079} {\path{arXiv:2409.03079}}.

\bibitem{CM2024-onesync}
E.~Carson and Y.~Ma.
\newblock A stable one-synchronization variant of reorthogonalized block classical {Gram--Schmidt}, 2024.
\newblock \href {https://arxiv.org/abs/2411.07077} {\path{arXiv:2411.07077}}.

\bibitem{trilinos2005}
M.~A. Heroux, R.~A. Bartlett, V.~E. Howle, R.~J. Hoekstra, J.~J. Hu, T.~G. Kolda, R.~B. Lehoucq, K.~R. Long, R.~P. Pawlowski, E.~T. Phipps, A.~G. Salinger, H.~K. Thornquist, R.~S. Tuminaro, J.~M. Willenbring, A.~Williams, and K.~S. Stanley.
\newblock An overview of the trilinos project.
\newblock {\em ACM Trans.\ Math.\ Software}, 31(3):397--423, 2005.
\newblock \href {https://doi.org/10.1145/1089014.1089021} {\path{doi:10.1145/1089014.1089021}}.

\bibitem{YTHBSE2020}
I.~Yamazaki, S.~Thomas, M.~Hoemmen, E.~G. Boman, K.~{\'S}wirydowicz, and J.~J. Elliott.
\newblock Low-synchronization orthogonalization schemes for s-step and pipelined {{Krylov}} solvers in {{Trilinos}}.
\newblock In {\em Proceedings of the 2020 {{SIAM Conference}} on {{Parallel Processing}} for {{Scientific Computing}} ({{PP}})}, pages 118--128, 2020.
\newblock \href {https://doi.org/10.1137/1.9781611976137.11} {\path{doi:10.1137/1.9781611976137.11}}.

\end{thebibliography}

\end{document}